\documentclass[aps,prl,showpacs,floatfix,twocolumn,amsmath,amssymb,preprintnumbers]{revtex4-1}
\usepackage{mathrsfs}
\usepackage[figuresright]{rotating}
\usepackage{amsmath}
\usepackage{amssymb}
\usepackage{graphicx}
\usepackage{color}
\usepackage{dcolumn}
\usepackage{bm}
\usepackage[breaklinks=true,colorlinks=true,linkcolor=blue,urlcolor=blue,citecolor=blue]{hyperref}
\UseRawInputEncoding   
\usepackage{soul}

\makeatletter
\newcommand{\rmnum}[1]{\romannumeral #1}
\newcommand{\Rmnum}[1]{\expandafter\@slowromancap\romannumeral #1@}
\makeatother

\begin{document}

\title{$^{27}$Al NMR study of the magnetic Weyl semimetal CeAlGe}

\author{Zhuo Wang$^{1}$}
\author{Xiaobo He$^{1}$}
\author{Fangjun Lu$^{1}$}
\author{Hai Zeng$^{1}$}
\author{Shuo Zou$^{1}$}
\author{Xiao-Xiao Zhang$^{1}$}
\author{Yongkang Luo$^{1}$}
\email[]{mpzslyk@gmail.com}
\address{$^1$Wuhan National High Magnetic Field Center and School of Physics, Huazhong University of Science and Technology, Wuhan 430074, China.}

\date{\today}

\begin{abstract}

Motivated by the recent observations of electronic correlation effect [M. Corasaniti \textit{et al}., Phys. Rev. B \textbf{104}, L121112 (2021)] and topology-stabilized magnetic fluctuations [N. Drucker \textit{et al}., Nat. Commun. \textbf{14}, 5182 (2023)] in the noncentrosymmetric magnetic Weyl semimetal candidate CeAlGe, we performed systematic studies on the local static and dynamic spin susceptibilities by $^{27}$Al nuclear magnetic resonance. Due to the large spin susceptibility from Ce-$4f$ electrons, the theoretically predicted responses from Weyl fermions are overwhelmed. A Knight-shift anomaly is observed below $T^*\sim50$ K, a signature of the onset of coherent Kondo coupling. In addition, an anomalous peak is found in $1/T_1T$ near 15 K, well above the magnetic ordering temperature $T_N \approx 5$ K, which probably is a consequence of topology-stabilized magnetic fluctuations. These results highlight the interplay among
electronic correlation, magnetism and band topology in this family of Kondo Weyl semimetals.

\end{abstract}


\maketitle


Recently, band-topology-mediated magnetism in Weyl semimetals has inspired fast growing interest for both fundamental research and potential applications \cite{DruckerZeroMR,Gaudet-NdAlSiNM2021,Yao-SmAlSiPRX2023}. Weyl semimetals are a class of topological quantum material whose properties are mainly determined by Weyl fermions, a kind of massless chiral quasiparticle that can be viewed as a ``half" of a Dirac fermion \cite{Weyl-Physik1929, Armitage-RMP2018}. Principally, a Weyl semimetal can be realized in materials where either spatial-inversion (SI) or time-reversal (TR) symmetry is broken \cite{WengTaAsSI,huangTaAsFerimarcs,HuangnegativeMR,MoraliFerimarcsCoSnS}. Unique electronic states and novel physical properties such as Fermi arcs \cite{xuScienceFerimarcsTaAs,LvPRLFerimarcsTaAs,yangTaAsFerimarcs,huangTaAsFerimarcs}, large magnetoresistance and high carrier mobility \cite{Ali-WTe2XMR,Zhang-TaAsMR}, non-trivial Berry phase \cite{LuoY-NbAsSdH,HuJ-TaPSdH}, chiral anomaly \cite{HuangnegativeMR}, and anomalous Hall effect \cite{BurkovAHE,wangAHE} have been ubiquitously seen in Weyl semimetals, endowing them with immense potential applicability in electronic and spintronic devices.

$R$Al$X$ ($R$ = Rare earth, $X$ = Si, Ge) represents a family of special Weyl semimetals in that both SI and TR symmetries are broken simultaneously \cite{ChangRAlGefamily}. Type-II Weyl points emerge in the non-magnetic LaAl$X$ that crystallize in the noncentrosymmetric $I4_{1}md$ (No. 109) structure, and they are further stabilized by the magnetic $R$ ions. Of prime interest are the members $R=$ Ce where the $4f^1$ electronic structure sets a versatile platform to explore the interplay among topology, magnetism and electronic correlation. Our previous resistivity and AC calorimetry measurements revealed pressure-enhanced antiferromagnetic (AFM) transition in CeAlGe \cite{HeCeAlGeDWs}; the similar trend was also found in CeAlSi \cite{PivaLTHE}. This places CeAl$X$ at a regime with relatively weak $c$-$f$ hybridization on Doniach's phase diagram \cite{Doniach}. Optical conductivity experiments, however, manifested a notable electronic correlation effect in CeAlGe below $\sim$ 100 K \cite{CorasanitiKondo100K}. A natural question concerns how such electronic correlation effect is facilitated in this low-carrier-density semimetal ($\sim$ 0.063 hole/f.u. \cite{HeCeAlGeDWs}, cf Nozi\`{e}res exhaustion problem \cite{Nozieres-EPJB1998,LuoY-CeNi2As2Pre}). On the other hand, a recent work by Drucker \textit{et al} reported the presence of locally correlated magnetism well above the thermodynamic magnetic transition temperature. More interestingly, the wavevector of this short-range order is consistent with the nesting condition of topological Weyl nodes \cite{DruckerZeroMR}, suggesting coupling between Weyl fermion and magnetism. \textit{Microscopic local} experiments are needed to confirm these intriguing features, and to unveil more details as well.

Herein, we employed $^{27}$Al nuclear magnetic resonance (NMR) measurements on CeAlGe, and both static and dynamic spin susceptibilities were investigated. We find both NMR shift and spin-lattice relaxation rate ($1/T_1$) in this compound are dominated by the Ce-$4f$ electrons, while the intrinsic responses from Weyl fermions are buried. Coherent Kondo scale $T^*\sim 50$ K is suggested by the Knight-shift anomaly, reaffirming a moderate electronic correlation. An anomalous peak is present in $1/T_1T$ near 15 K, lending further support for topology-stabilized magnetic fluctuations prior to magnetic ordering. Our work corroborates CeAlGe as a rare example of Kondo Weyl semimetal whose properties are determined by a combination of topology, magnetism and electronic correlation.


High-quality CeAlGe single crystals were grown by the Al-flux method as described elsewhere \cite{HeCeAlGeDWs}. The sample quality was verified by single crystalline X-ray diffraction and energy-dispersive X-ray spectroscopy. $^{27}$Al (gyromagnetic ratio $^{27}\gamma_n$ = 11.0943 MHz/T) NMR spectra were recorded in a stepped frequency sweep spin-echo method in the presence of a fixed external field $\mu_0{H}\approx 7$ T, $\mathbf{H}\parallel\mathbf{c}$. A small piece of aluminium foil was placed inside the sample coil, serving as a reference to the $^{27}$Al signal in CeAlGe. The value of $H$ was verified by the $^{63}$Cu shift in the sample coil. Spin-lattice relaxation rate was measured in a standard inversion recovery method on the central ($\frac{1}{2}\leftrightarrow-\frac{1}{2}$) transition, and $T_1$ was extracted by fitting the recovery curve to
\begin{equation}
\begin{aligned}
M(t)=M(\infty)\{&1-\frac{2f}{315}[9\exp{(\frac{-t}{T_{1}})}+56\exp{(\frac{-6t}{T_{1}}}) \\
&+250\exp{(\frac{-15t}{T_{1}})}]\},
\end{aligned}
\label{Eq1}
\end{equation}
where $M(\infty)$, $f$ and $T_1$ are fitting parameters. The fitting results at selected temperatures are provided in \textbf{Supplemental Material, SM}\cite{SM}. Magnetic susceptibility was measured in a magnetic property measurement system (MPMS, Quantum Design) equipped with a vibrating sample magnetometer (VSM).


\begin{figure}[!ht]
\vspace*{-5pt}
\hspace*{-13pt}
\includegraphics[width=9.5cm]{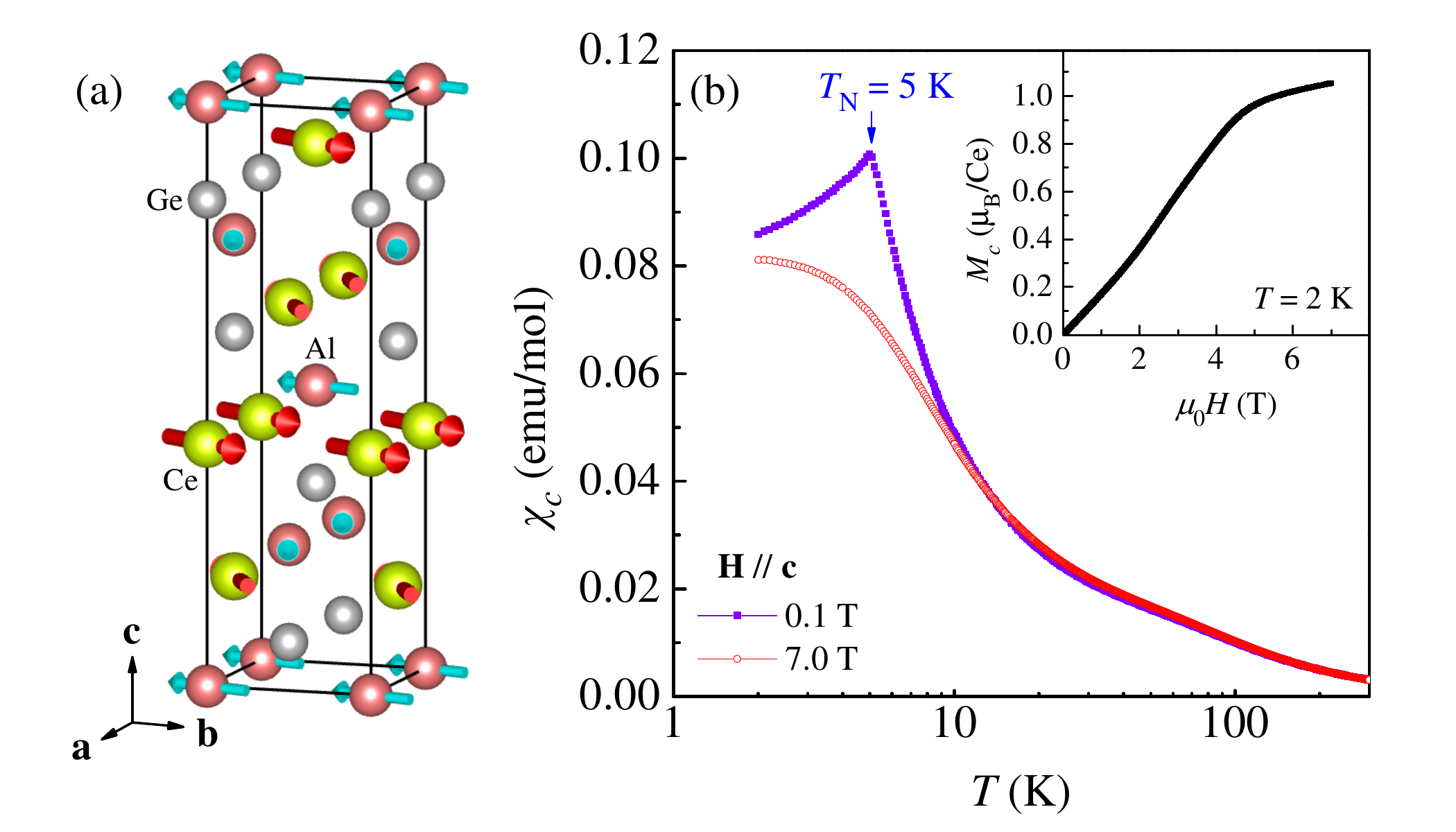}
\vspace*{-15pt}
\caption{(a) Crystal structure of CeAlGe and local environment of Al. The red arrows denote Ce moments. The cyan arrows indicate the dipolar field at Al sites calculated for the magnetic structure proposed in \cite{suzukiScienceDWs}. (b) Temperature dependence of magnetic susceptibility measured at $\mu_0H$ = 0.1 T (violet) and 7.0 T (red). The inset shows the magnetization isotherm $M(H)$ at 2 K.}
\label{Fig1}
\end{figure}

Figure \ref{Fig1}(b) shows the temperature dependence of magnetic susceptibility ($\chi$) measured with $\mathbf{H}\parallel\mathbf{c}$. At high temperature, $\chi_c(T)$ conforms to a Curie-Weiss behavior with $\Theta_W \approx 37.5$ K. An AFM-like peak is observed near 5 K under an external field of 0.1 T (violet). The peak is gone as field is increased to 7 T, and $\chi_c(T)$ tends to level off at low temperature. This suggests that a field of 7 T polarizes the Ce moments, which is confirmed by the isothermal magnetization $M(H)$ as shown in the inset. Previous anisotropic magnetic susceptibility \cite{Puphal-CeAlGe-THE} and neutron scattering \cite{suzukiScienceDWs} experiments on CeAlGe have revealed an easy-plane $Fd'd2'$ configuration in the ordered state, with the order parameters $\mathbf{m}_A$ = ($m_x$, $m_y$, 0) for sublattice A and $\mathbf{m}_B$ = ($-m_y$, $-m_x$, 0) for sublattice B; $\mathbf{m}_A$ and $\mathbf{m}_B$ are noncollinear. A schematic magnetic structure is provided in Fig.~\ref{Fig1}(a). By adopting this configuration, we calculate the dipolar field on the Al sites exerted by the Ce moments, as depicted by the cyan arrows. The obtained dipolar fields for the two sublattices are: $\mu_0\mathbf{H}^{dip}_A=(-0.0129,~-0.0389,~ 0)$ T and $\mu_0\mathbf{H}^{dip}_B=(0.0389,~0.0129,~0)$ T.

\begin{figure}[!htp]
\vspace{4pt}
\hspace{0pt}
\includegraphics[width=8.5cm]{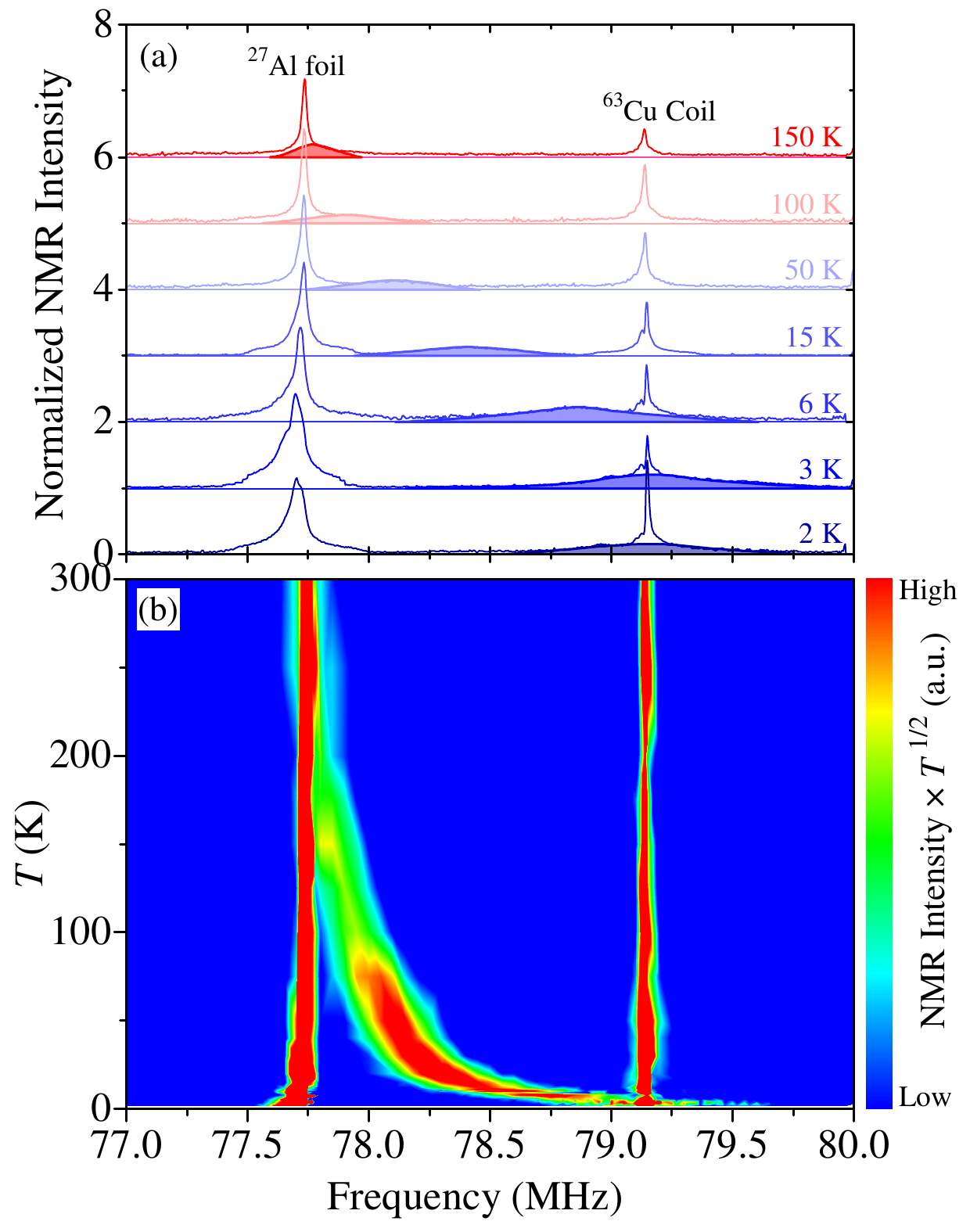}
\vspace{-10pt}
\caption{$^{27}$Al frequency-sweep NMR spectra of CeAlGe measured under $\mu_0H\approx$ 7 T, $\mathbf{H}\parallel\mathbf{c}$. (a) Normalized spectra at representative temperatures. The curves are vertically offset for clarity. The peaks for $^{27}$Al in CeAlGe are filled with colors. (b) Contour plot of NMR spectra for all the measured temperatures 2-300 K.  }
\label{Fig2}
\end{figure}

The $^{27}$Al NMR spectra of CeAlGe under $\mu_0 H \simeq 7$ T are shown in Fig.~\ref{Fig2}. Representative spectra at selected temperatures, 2, 3, 6, 15, 50, 100 and 150 K, are provided in Fig.~\ref{Fig2}(a). Three peaks can be recognized in the frequency window interested, assigned to $^{27}$Al $\frac{1}{2}\leftrightarrow-\frac{1}{2}$ in foil, $^{27}$Al $\frac{1}{2}\leftrightarrow-\frac{1}{2}$ in CeAlGe, and $^{63}$Cu $\frac{1}{2}\leftrightarrow-\frac{1}{2}$ in coil, respectively. Both $^{27}$Al-foil and $^{63}$Cu-coil peaks are essentially temperature independent, as expected. The $^{27}$Al in sample overlaps with that in aluminium foil at 300 K, and upon cooling, the peak gradually moves to the right, and meanwhile the linewidth broadens substantially, implying an enhanced magnetic correlation. The evolution of the resonance peaks can be well seen in a false-color contour plot shown in Fig.~\ref{Fig2}(b).

\begin{figure}[!htp]
\vspace{-0pt}
\hspace{0pt}
\includegraphics[width=8.5cm]{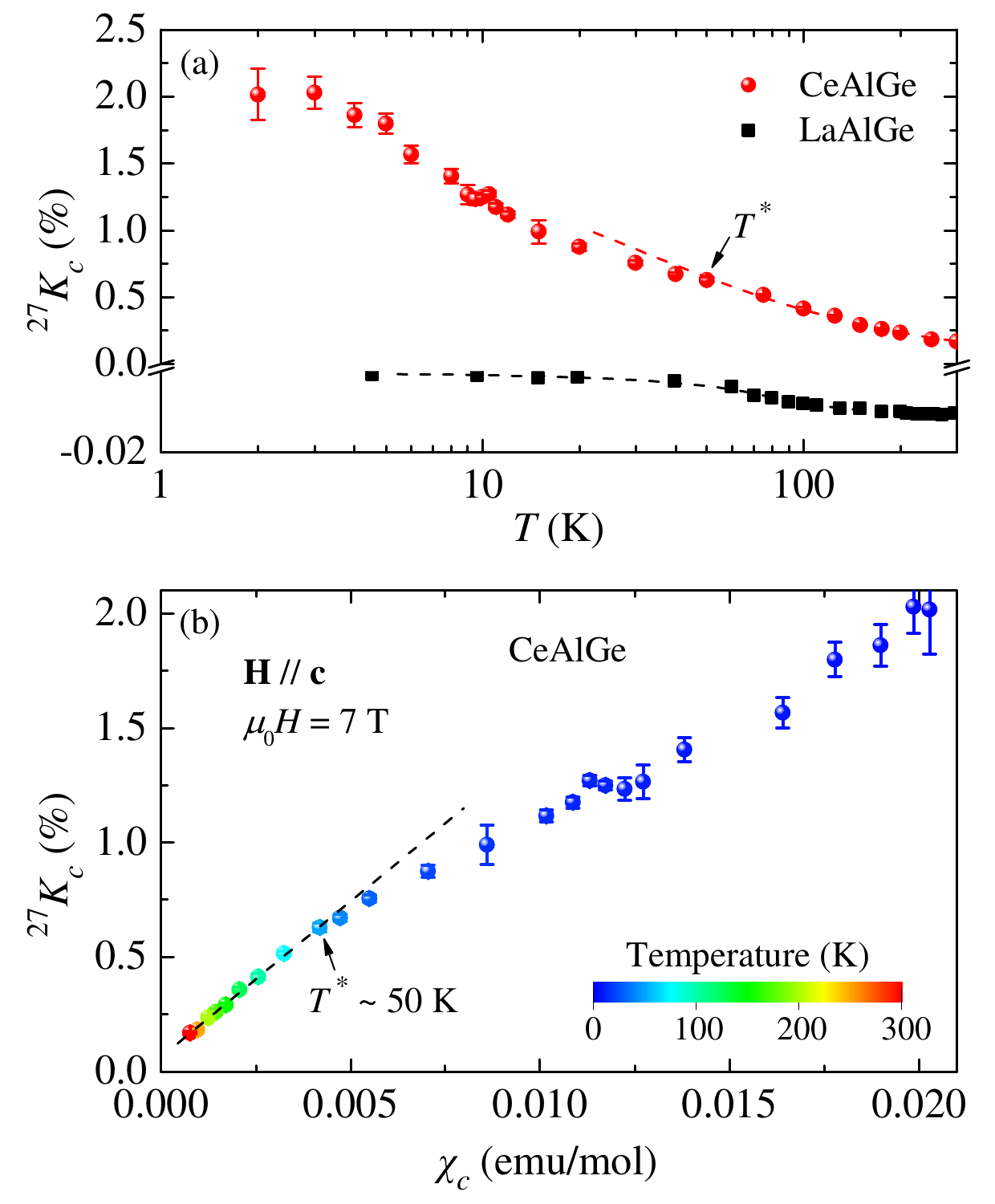}
\vspace{-15pt}
\caption{(a) Temperature dependence of $^{27}$Al NMR shift in CeAlGe. The data of LaAlGe are reproduced from Ref.~\cite{TayPrAlGeNMR}. The dashed lines denote the  fittings to Eq.~(\ref{Eq.2}) for LaAlGe and Curie-Weiss formula for CeAlGe, respectively. 
(b) The Clogston-Jaccarino plot $^{27}K$ vs. $\chi$ with $T$ as an implicit parameter. The Knight-shift anomaly near 50 K is ascribed to Kondo coherence.}
\label{Fig3}
\end{figure}

NMR shift ($K$) can be extracted from the $^{27}$Al spectra in a second-order quadrupolar effect approximation \cite{RybickiKshiftformula}, and the results are shown in Fig.~\ref{Fig3}(a) as a function of temperature. Multimodal Gaussian fitting was processed for temperatures above 150 K, seeing \textbf{SM} \cite{SM}. For comparison, the results of LaAlGe (data from Ref.~\cite{TayPrAlGeNMR}) are also shown. For Weyl semimetals, earlier theoretical works predicted that NMR shift \cite{TayPrAlGeNMR,OkvatovityT1model2019}:
\begin{equation}
K(\mu, T) \approx \frac{\mu_0e}{4\pi^2\hbar}[\frac{g\mu_B}
{\hbar v_F}\mu-\frac{e v_F}{3}\ln(\frac{W}{max\{|\mu|, k_B T]\}})],
\label{Eq.2}
\end{equation}
where $\mu$ is chemical potential, $v_F$ is Fermi velocity, $W$ is a sharp high-energy cutoff regularizing the
theory, while $\hbar$, $k_B$, $\mu_B$, $\mu_0$ and $e$ are physical constants of conventional meanings. This formula was found to fit the nonmagnetic LaAlGe rather satisfactorily, seeing the dashed line in Fig.~\ref{Fig3}(a) \cite{TayPrAlGeNMR}. Unlike LaAlGe whose NMR shift is both negative and small, $^{27}K_c$ of CeAlGe is positive all through the temperature range 2-300 K, and the magnitude is also much larger. At 300 K, $^{27}K_c \approx 0.17\%$. As temperature decreases, $^{27}K_c$ increases monotonically, roughly following the Curie-Weiss formula. Korringa region - where $K$ is independent of $T$ at low temperature \cite{WangTaAsNMR,PapawassiliouWTe2NMR} - is not clear in CeAlGe. All these imply that the NMR shift in CeAlGe is dominated by the spin susceptibility from Ce-$4f$ electrons, whereas Weyl fermion contributes only a little, if any.

In metals, NMR shift usually decomposes into several components,
\begin{equation}
K(T)=K^o+K^s(T),
\label{Eq.3}
\end{equation}
where $K^o$ is orbital shift that is nearly temperature independent, and $K^s(T)$ is spin (Knight) shift and is related to the spin susceptibility by $K^s(T)=A_{hf}\chi(T)$. The ratio $A_{hf}$ is the hyperfine coupling constant. This linear relationship is better expressed by the so-called Clogston-Jaccarino plot $K$ vs. $\chi$ with $T$ as an implicit parameter, as shown in Fig.~\ref{Fig3}(b). We find the Clogston-Jaccarino relation is well obeyed for $T>50$ K with the slope $A_{hf}=0.79(2)$ T/$\mu_B$ but becomes gradually deviated below 50 K. In other words, $A_{hf}$ becomes temperature dependent for $T<50$ K. In Kondo lattices, such a violation is usually named as Knight-shift anomaly \cite{Jiang-KAnomaly,Curro_RPP2016}. We note that the Knight-shift anomaly observed in CeAlGe can not be attributed to CEF splitting of Ce$^{3+}$ $j=5/2$ sextet \cite{Ohama-CeCu2Si2KAno}, because a previous inelastic neutron scattering experiment revealed that the energy gap between the ground and first excited doublets is about 100 meV \cite{DruckerZeroMR}, orders of magnitude higher than 50 K. An alternative explanation to the Knight-shift anomaly is due to the onset of Kondo coherence. A phenomenological understanding can be provided by a two-fluid model which states that two different
fluids - an itinerant heavy-electron fluid and a Kondo impurity fluid - coexist below the Kondo coherence temperature $T^*$ \cite{Nakatsuji-2Fluid,Yang-PRL2008,Yang-PNAS2012,Shirer-PNAS2012,Curro2004Twofluid}. These two fluids are of different hyperfine coupling constants, and upon cooling from $T^*$, the weight of heavy-electron fluid (that is a consequence of $c$-$f$ hybridization) gradually increases at the expense of Kondo impurity fluid, and therefore the $K$ vs. $\chi$ plot loses linearity. For above $T^*$, only the Kondo impurity fluid survives, and the Clogston-Jaccarino relation restores. In this sense, our $^{27}$Al NMR experiment probably gives an estimate of the coherent Kondo scale, $T^*\sim 50$ K, in this Kondo Weyl semimetal. It should be pointed out that a recent optical conductivity measurements on CeAlGe also manifested a reduction of Fermi velocity and hence an enhancement of quasiparticle effective mass for temperatures below $\sim 100$ K \cite{CorasanitiKondo100K}, qualitatively in agreement with our $T^*$. Combining with NMR and optical conductivity works, it is rather likely that electronic correlation arising from $c$-$f$ renormalization appears at low temperature in this semimetallic Kondo lattice \cite{LuoY-CeNi2As2Pre,Pan-CeBiRUS}.

We should mention that at low temperature below $T_N$, $^{27}K_{c}(T)$ tends to saturate at 2\%, following the similar tendency of $\chi_{c}(T)$ under the same magnetic field, Fig.~\ref{Fig1}(b). This indicates that the out-of-plane component of the internal field at the Al site is $\sim 0.14$ T, about 3.5 times larger than the dipolar field we calculated. That is to say, the internal field is more dominated by a transferred hyperfine field arising from orbital hybridization, as is the case in other cerium-based heavy-fermion compounds (see CeNiAsO for instance \cite{LuFangjunCeNiAsO}). It is also interesting to note that $^{27}K_c$ of CeAlGe, although orders of magnitude larger than in LaAlGe, is much smaller than that of PrAlGe ($\sim 20\%$ at 6 K \cite{TayPrAlGeNMR}), probably because the hyperfine coupling is way stronger in the latter.

\begin{figure}[!ht]
\vspace{-0pt}
\hspace{-0pt}
\includegraphics[width=8.5cm]{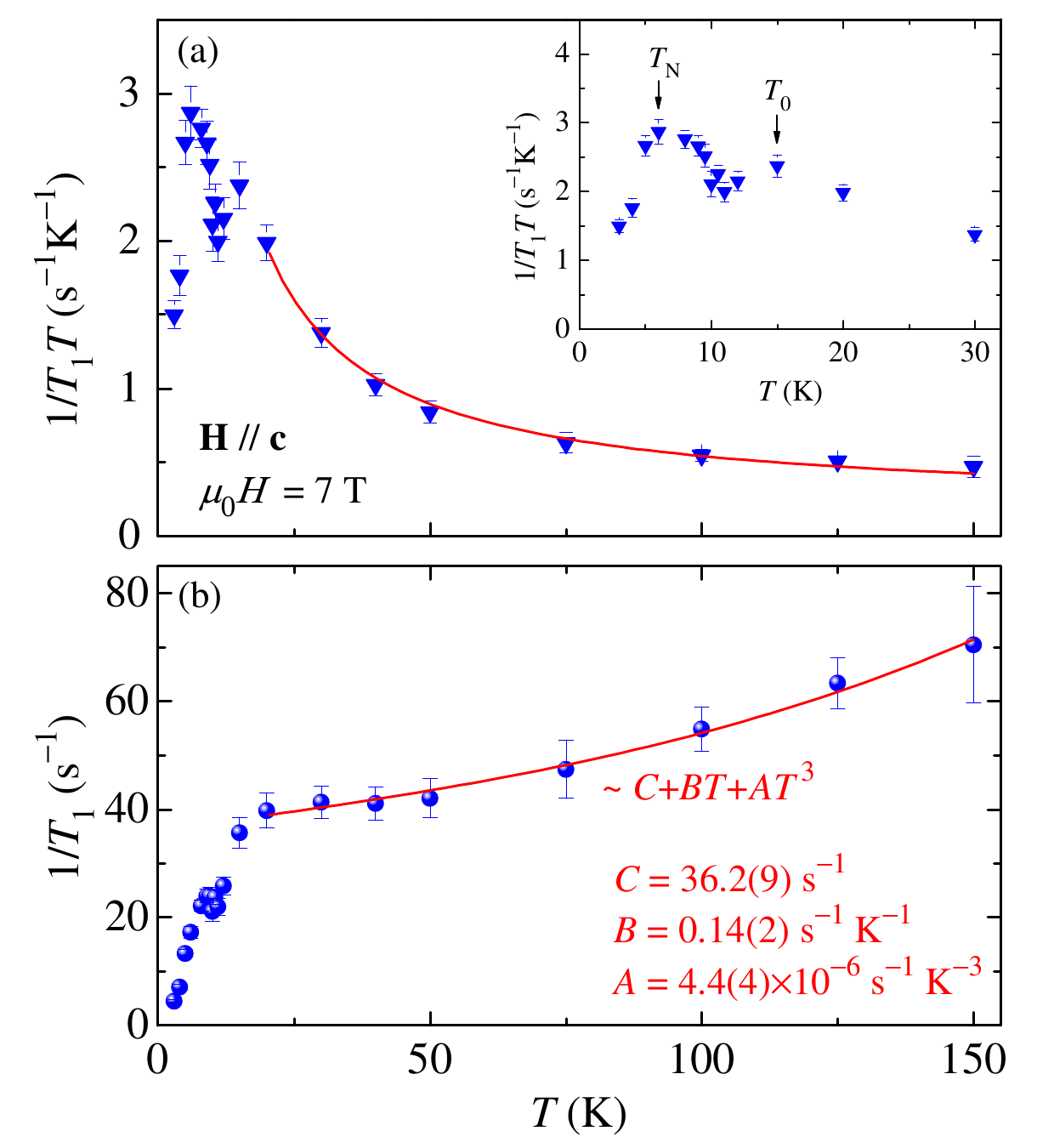}
\vspace{-15pt}
\caption{Spin-lattice relaxation rate of CeAlGe. (a) $T$ dependence of $1/T_1T$. The solid line is a Curie-Weiss fitting. Inset, a zoom-in view to show the features around $T_N$ and $T_0$. (b) $1/T_1$ as a function of $T$. The high-$T$ part of $1/T_1$ can be fitted to a $C+BT+AT^3$ law. }
\label{Fig4}
\end{figure}

Spin dynamics of CeAlGe is investigated by spin-lattice relaxation rate measurements (Fig.~\ref{Fig4}). For weakly-interacted metals like Cu, $1/T_1T$ probes $N^2(E_F)$, where $N(E_F)$ is the density of states at the Fermi level \cite{Abragam,Slichter}. For Weyl semimetals, earlier works predicted that $1/T_1T$ remains constant at low temperatures ($k_BT\ll\mu$) and changes into a $T^2\ln T$ dependence at high temperature ($k_BT\gg\mu$) \cite{OkvatovityT1model2016,OkvatovityT1model2019}, as expressed by
\begin{equation}
\begin{aligned}
\frac{1}{T_1T}=&\frac{52.7\pi\mu_0^2\gamma_n^2e^2k_B}{(2\pi)^6v_F^2\hbar} \\
&\times
\left\{\begin{aligned}
&\frac{\pi^2k_B^2T^2}{3\hbar^2}\ln{(\frac{4k_B T}{\hbar \omega_0})},  &k_BT\gg\mu,\\
&(\frac{\mu}{\hbar})^2\ln{(\frac{2\mu}{\hbar \omega_0})}, &k_BT\ll\mu,
\end{aligned}
\ \right.
\label{Eq.4}
\end{aligned}
\end{equation}
where $\omega_0$ is the nuclear Larmor frequency. Experimentally, a $T^2$-law was found to fit both Ta(P, As) \cite{WangTaAsNMR,YasuokaTaPNMR} and WTe$_2$ \cite{PapawassiliouWTe2NMR} well. For CeAlGe, however, due to the local moments of Ce $4f$ electrons, a dominant Curie-Weiss term is expected, as is shown in Fig.~\ref{Fig4}(a). We note that $1/T_1$ for above 150 K is hard to obtain in our experiment, because the sample signal well overlaps with that from the aluminium foil. In order to dig out more information, we plot $1/T_1$ as a function of $T$ in Fig.~\ref{Fig4}(b). $1/T_1$ seems essentially constant between 20 and 50 K as expected for a Curie-Weiss behavior \cite{Kohori-CeIn31999,Sakai-CePt2In72014,LuFangjunCeNiAsO}. To our surprise, $1/T_1$ turns up obviously above 50 K. We argue that the increase of $1/T_1$ at high temperature may have two origins: (\rmnum{1}) the Fermi-liquid term with $1/T_1 \propto T$, and (\rmnum{2}) the Weyl-fermion term with $1/T_1 \propto T^3$. Indeed, a \textit{superlinear} $C+BT+AT^3$ formula can fit the $1/T_1$ data nicely, seeing the red line in Fig.~\ref{Fig4}(b). This suggests that scatterings with Weyl fermion do contribute a channel to spin-lattice relaxation, albeit to a small extent.

At low temperature, a pronounced peak is observed in $1/T_1T$ near $T_N$. Since $1/T_1T$ probes the transverse fluctuation field, although the application of a longitudinal field has nearly polarized the Ce moments, it is conceivable that transverse magnetic fluctuations can survive, and they are suppressed when the Ce moments are AFM ordered. To our interest, another hump centering around $T_0\sim 15$ K is visible in $1/T_1T$ [inset to Fig.~\ref{Fig4}(a)]. This suggests the proliferation of magnetic fluctuations prior to the thermodynamic transition, and is reminiscent of the recent work lead by N. Drucker \cite{DruckerZeroMR}; there electrical transport, thermal transport, resonant elastic X-ray scattering, and dilatometry measurements consistently indicated the presence of a short-range magnetic order with an incommensurate wavevector $\mathbf{q_m}=(0.384,~0.416,~0)$ at around 13 K. Note that $\mathbf{q_m}$ coincides with that of possible nesting between the type-I W$_3$ Weyl nodes located at ($\pm$0.2, $\pm$0.2, 0).

\begin{figure}[!ht]
\vspace{-20pt}
\hspace{-0pt}
\includegraphics[width=8.5cm]{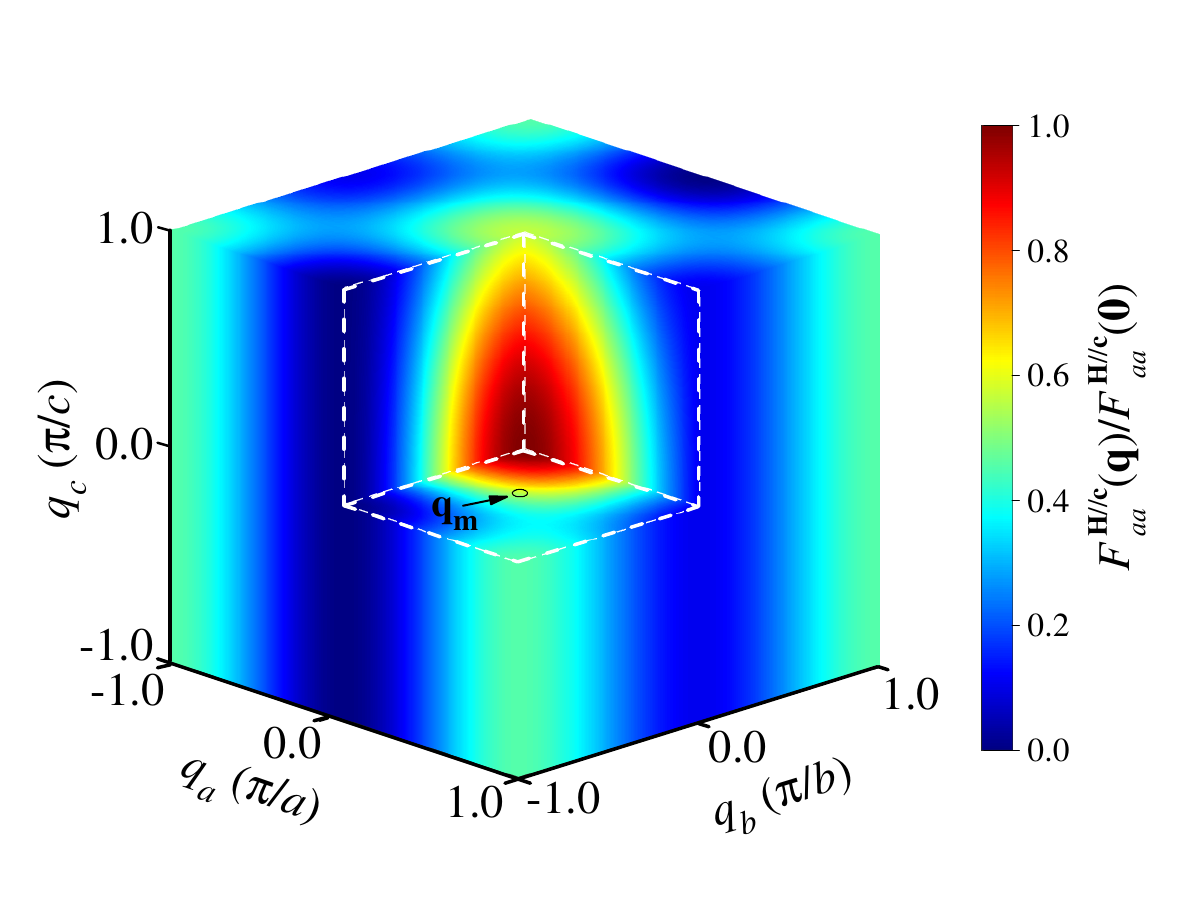}
\vspace{-25pt}
\caption{Distribution of hyperfine form factor $F^{\mathbf{H}\parallel\mathbf{c}}_{aa} (\mathbf{q})$ in momentum space. The results are presented after normalized by the value at $\mathbf{q}=\mathbf{0}$. }
\label{Fig5}
\end{figure}

To further discuss about the possibility of Weyl-fermion mediated magnetism, we calculated the hyperfine form factor [$F_{\alpha\beta}(\mathbf{q}$); $\alpha,\beta=a,b,c$] at the Al site in CeAlGe. According to Moriya's spin fluctuations theorem, $1/T_1T$ is connected to the dynamic spin susceptibility [$\chi(\mathbf{q},\omega)$] by \cite{Moriya-T1SF,Moriya1974}
\begin{equation}
(\frac{1}{T_1T})_{\mathbf{H}\parallel\mathbf{c}}=\lim_{\omega \rightarrow 0}\frac{\gamma_n^2 k_B}{N}\sum_{\mathbf{q}}F_{aa}^{\mathbf{H}\parallel\mathbf{c}}(\mathbf{q})\frac{\chi''_{aa}(\mathbf{q},\omega)}{\hbar \omega},
\label{Eq.5}
\end{equation}
\textit{i.e.}, $F_{\alpha\beta}(\mathbf{q})$ serves as a weight when summing the $\mathbf{q}$ dependent $\chi''$ into the $\mathbf{q}$-averaged $1/T_1T$. Note that in Eq.~(\ref{Eq.5}) we have already adopted the tetragonal symmetry of CeAlGe. More details about the calculations can be found in \textbf{SM} \cite{SM}, seeing also Refs.~\cite{SmeraldBaFe2As2NMR,ZhaoYFe2Ge2NMR}.
The calculated distribution of hyperfine form factor in momentum space, after normalized by its maximum at $\mathbf{q}=\mathbf{0}$, is presented in Fig.~\ref{Fig5}. The normalized hyperfine form factor decays from the center of Brillouin zone, and retains a large value 0.558 at $\mathbf{q_m}$, demonstrating that spin fluctuations caused by nesting between the Weyl nodes at ($\pm$0.2, $\pm$0.2, 0) can be captured by $^{27}$Al $1/T_1T$. A new question then may be put forward: why these topology-stabilized fluctuations finally do not condense into a long-range order, but instead, an AFM order with propagation vector $\sim$ (0.066, 0.066, 0) appears below $T_N$ \cite{DruckerZeroMR,Puphal-CeAlGe-THE}? The reason for this ``failed phase transition" invites more investigations in the future.


In conclusion, by $^{27}$Al NMR experiments on the magnetic Weyl semimetal CeAlGe, we convey two important messages: (\rmnum{1}) a Knight-shift anomaly is observed below $\sim$ 50 K, which gives a measure of the coherent Kondo scale and thus highlights the role played by electronic correlation in this semimetallic Kondo lattice; (\rmnum{2}) magnetic fluctuations well above $T_N$ are detected by spin-lattice relaxation, which probably are a consequence of topology-stabilized short-range ordering. Our work, therefore, invokes further considerations about the interplay among electronic correlation, magnetism and band topology in this family of Kondo Weyl semimetals.


This work is supported by National Key R\&D Program of China (2023YFA1609600, 2022YFA1602602), Fundamental Research Funds for the Central Universities (YCJJ20230108), National Natural Science Foundation of China (U23A20580), the open research fund of Songshan Lake Materials Laboratory (2022SLABFN27), and Guangdong Basic and Applied Basic Research Foundation (2022B1515120020).


%

\newpage

\renewcommand{\thefigure}{S\arabic{figure}}
\renewcommand{\thetable}{S\arabic{table}}
\renewcommand{\theequation}{S\arabic{equation}}
\onecolumngrid

\newpage

\begin{center}
{\bf \large
{\it Supplemental Material:}\\
$^{27}$Al NMR study of the magnetic Weyl semimetal CeAlGe
}
\end{center}

\setcounter{table}{0}
\setcounter{figure}{0}
\setcounter{equation}{0}
\setcounter{page}{1}

\small
\begin{center}

Zhuo Wang$^{1}$,Xiaobo He$^{1}$,Fangjun Lu$^{1}$,Hai Zeng$^{1}$,Shuo Zou$^{1}$,Xiao-Xiao Zhang$^{1}$,Yongkang Luo$^{1*}$\email{mpzslyk@gmail.com}\\
$^1${\it Wuhan National High Magnetic Field Center and School of Physics, Huazhong University of Science and Technology, Wuhan 430074, China.}\\
\date{\today}
\end{center}
\normalsize
\vspace*{0pt}

In this \textbf{Supplemental Material (SM)}, we provide additional results that will further support the discussions and conclusion in the main text, including multimodal Gaussian fitting of the NMR spectrum, $T_1$ fitting of the recovery curves, and the details about the calculation of $^{27}$Al hyperfine form factor.

\section{SM I. M\lowercase{ultimodal} G\lowercase{aussian fitting of the} NMR \lowercase{spectrum}}

\begin{figure*}[!htp]
\vspace*{-10pt}
\includegraphics[width=10cm]{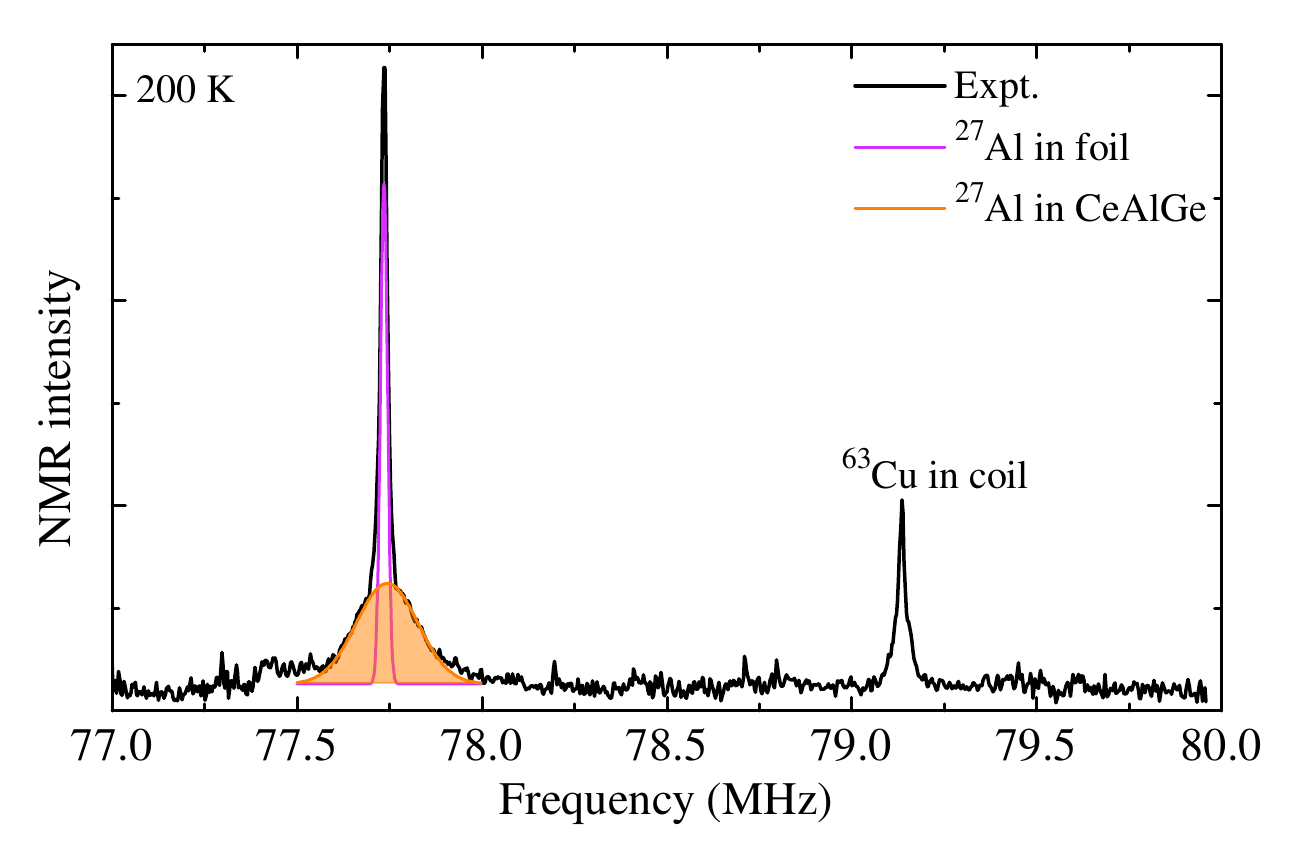}
\vspace*{-10pt}
\caption{\label{FigS1} Multimodal Gaussian fitting of the $^{27}$Al NMR spectrum at 200 K. The broad peak filled with orange is the $^{27}$Al $\frac{1}{2}\leftrightarrow-\frac{1}{2}$ in CeAlGe, the magenta peak is the $^{27}$Al signal in aluminum foil, and the black line is the experimental curve. }
\end{figure*}

By a stepped frequency sweep spin-echo method, the $^{27}$Al signals from CeAlGe and aluminum foil were recorded simultaneously, and the latter was used as a reference when calculating the NMR shift $K$. However, for temperatures above 150 K, these signals become well overlapped with each other, so a multimodal Gaussian fitting is needed. As an example, in Fig.~\ref{FigS1}, we display this fitting for the spectrum at 200 K.

\section{SM II. S\lowercase{pin-lattice relaxation rate fit}}

\begin{figure*}[!htp]
\vspace*{10pt}
\includegraphics[width=18cm]{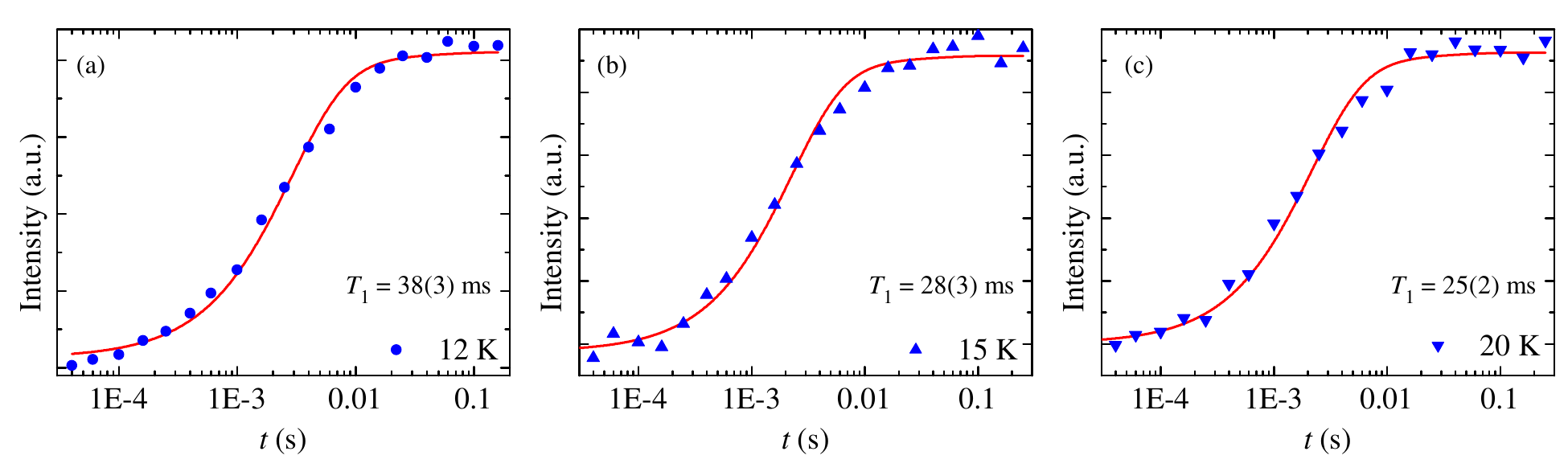}
\vspace*{-15pt}
\caption{\label{FigS2} Inversion recovery $M(t)$ curves of $^{27}$Al $\frac{1}{2} \leftrightarrow -\frac{1}{2}$ resonance in CeAlGe. The red solid lines are the fittings to Eq.~(\ref{Eq.S1}) to extract $T_1$. (a) 12 K; (b) 15 K; (c) 20 K. }
\end{figure*}

Spin-lattice relaxation rate was measured in a standard inversion recovery method on the central ($\frac{1}{2}\leftrightarrow-\frac{1}{2}$) transition, and $T_1$ was extracted by fitting the recovery curve $M(t)$ to
\begin{equation}
\begin{aligned}
M(t)=M(\infty)\{&1-\frac{2f}{315}[9\exp{(\frac{-t}{T_{1}})}+56\exp{(\frac{-6t}{T_{1}}}) \\
&+250\exp{(\frac{-15t}{T_{1}})}]\},
\end{aligned}
\label{Eq.S1}
\end{equation}
where $M(\infty)$, $f$ and $T_1$ are fitting parameters. The recovery curves and their fittings to Eq.~(\ref{Eq.S1}) at 12, 15 and 20 K are provided as examples, cf Fig.~\ref{FigS2}.

\newpage

\section{SM III. H\lowercase{yperfine form factor calculation}}

In this part, we provide the detail about the calculation of hyperfine form factor at the Al site in CeAlGe. According to Moriya's spin fluctuations theorem, $1/T_1T$ is connected to the dynamic spin susceptibility [$\chi(\mathbf{q},\omega)$] \cite{Moriya-T1SF,Moriya1974}. For $\mathbf{H} \parallel \mathbf{c}$ in tetragonal symmetry, $1/T_1T$ can be written as
\begin{equation}
(\frac{1}{T_1T})_{\mathbf{H}\parallel\mathbf{c}}=\lim_{\omega \rightarrow 0}\frac{\gamma_n^2 k_B}{N}\sum_{\mathbf{q}}F_{aa}^{\mathbf{H}\parallel\mathbf{c}}(\mathbf{q})\frac{\chi''_{aa}(\mathbf{q},\omega)}{\hbar \omega},
\label{Eq.S2}
\end{equation}
where $k_B$ is Boltzmann's constant, $\hbar$ is reduced Planck constant, $N$ is the number of atoms in a primitive cell, $\chi''_{\alpha\beta}(\omega,\mathbf{q})$ is the imaginary part of the dynamic spin susceptibility at wave vector $\mathbf{q}$ and frequency $\omega$ ($\alpha, \beta$ = a, b, c),, and $F_{\alpha\beta}(\mathbf{q})$ serves as a weight when summing the $\mathbf{q}$ dependent $\chi''$ into the $\mathbf{q}$-averaged $1/T_1T$.

\begin{figure*}[!htp]
\vspace*{-0pt}
\includegraphics[width=16cm]{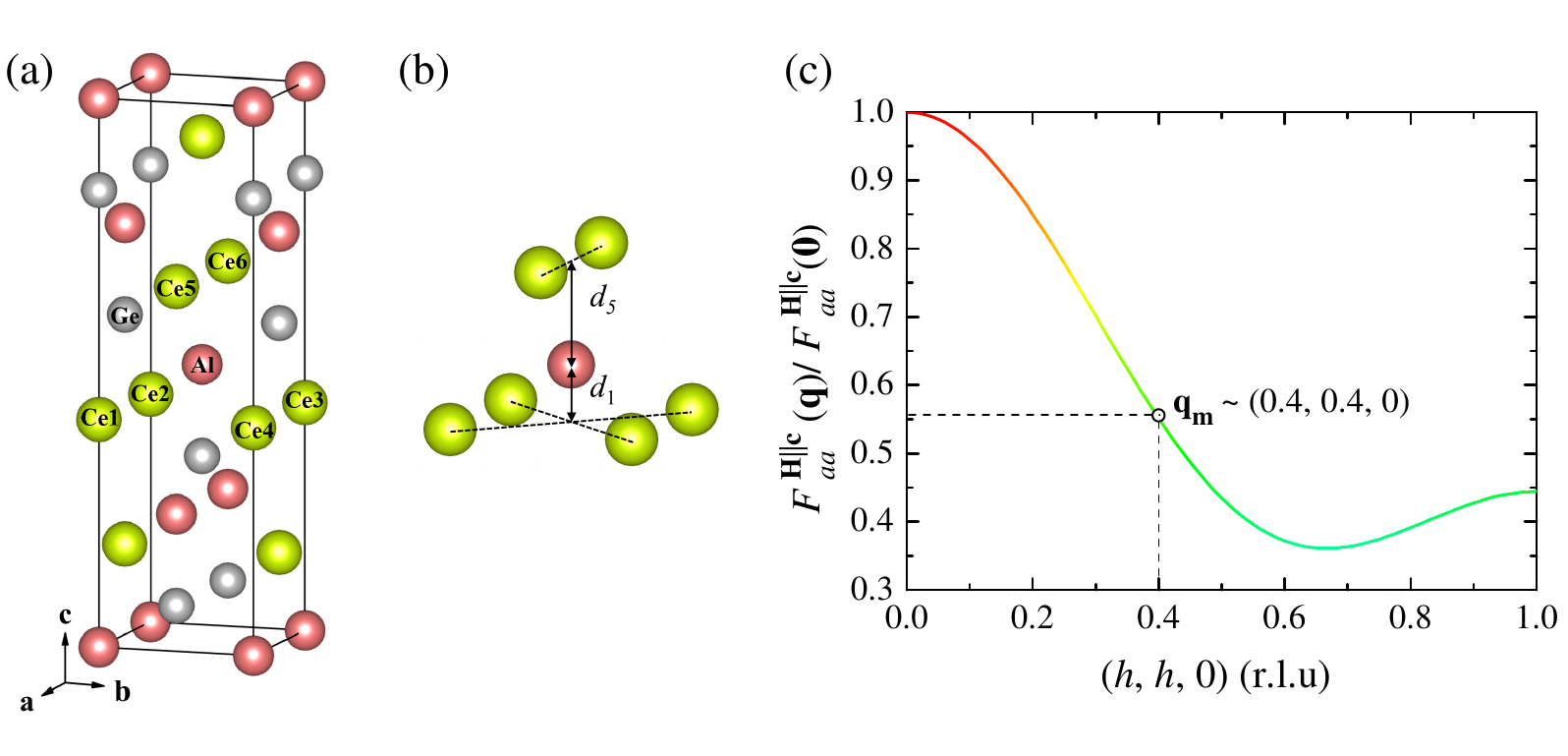}
\vspace*{-10pt}
\caption{The hyperfine form factor calculation of $^{27}$Al in CeAlGe. (a) Crystal structure of CeAlGe. Al is surrounded by 2 nearest neighbors Ce atoms (Ce5 and Ce6, $d_{\text{Ce-Al}}$=3.234 \AA) and 4 next-nearest neighbors Ce (Ce1 to Ce4, $d_{\text{Ce-Al}}$=3.275 \AA). (b) $d_1=0.085c$ and $d_5=0.165c$ represent the components of Ce1 and Ce5 to Al in the z direction, respectively. (c) The normalized hyperfine form factor as a function of $\mathbf{q}$ along ($h$, $h$, 0). Note that $\mathbf{q_m}$ locates near (0.4, 0.4, 0) in this plot.}
\label{FigS3}
\end{figure*}

$F_{\alpha\beta}(\mathbf{q})$ can be calculated via \cite{SmeraldBaFe2As2NMR,ZhaoYFe2Ge2NMR}
\begin{equation}
	F_{\alpha\beta}(\mathbf{q})=\mathbb{A}_{x\alpha}(\mathbf{q})\mathbb{A}_{x\beta}(-\mathbf{q})+\mathbb{A}_{y\alpha}(\mathbf{q})\mathbb{A}_{y\beta}(-\mathbf{q}),
	\label{Eq.S3}
\end{equation}
where ($x$,~$y$,~$z$) represents the Cartesian coordinate system where $z$-axis is parallel to magnetic field. The matrix element of the transferred hyperfine coupling tensor is given by $\mathbb{A}_{\alpha\beta}(\mathbf{q})$=$\sum_{i}\mathbf{A}^i_{\alpha\beta}(\mathbf{q})e^{i\mathbf{q}\cdot\mathbf{r}_i}$. The sum should be over all Ce sites around Al. However, the hyperfine coupling rapidly decays as distance increasing. To simplify calculations, only the six nearest-neighbor Ce atoms were considered, two of which (Ce5 and Ce6) have a distance of 3.234 \AA~ and four of which (Ce1 to Ce4) have a distance of 3.275 \AA. According to the labeling of the Ce moments shown in Fig.~\ref{FigS3}(a), the transferred hyperfine coupling tensors for Ce1 and Ce5 sites can be written as
\begin{equation}
		\mathbf{A}^1=
		\left [ \begin{matrix}
			A_{aa}& A_{ab} &A_{ac}\\
			A_{ba}& A_{bb} &A_{bc}\\
			A_{ca}& A_{cb} &A_{cc}\\
		\end{matrix} \right ],
            \mathbf{A}^5=
		\left [\begin{matrix}
			B_{aa}& B_{ab} &B_{ac}\\
			B_{ba}& B_{bb} &B_{bc}\\
			B_{ca}& B_{cb} &B_{cc}\\
		\end{matrix} \right ].
  \label{Eq.S4}
\end{equation}
Sequentially, $\mathbf{A}^2, \mathbf{A}^3,\mathbf{A}^4$ and $\mathbf{A}^6$ can be deduced by symmetry
\begin{equation}
		\mathbf{A}^2=
		\left [\begin{matrix}
			A_{aa}& -A_{ab} &-A_{ac}\\
			-A_{ba}& A_{bb} &A_{bc}\\
			-A_{ca}& A_{cb} &A_{cc}\\
		\end{matrix} \right ],
		\mathbf{A}^3=
		\left [\begin{matrix}
			A_{aa}& A_{ab} &-A_{ac}\\
			A_{ba}& A_{bb} &-A_{bc}\\
			-A_{ca}& -A_{cb} &A_{cc}\\
		\end{matrix} \right ],
		\mathbf{A}^4=
		\left [\begin{matrix}
			A_{aa}& -A_{ab} &A_{ac}\\
			-A_{ba}& A_{bb} &-A_{bc}\\
			A_{ca}& -A_{cb} &A_{cc}\\
		\end{matrix} \right ],
            \mathbf{A}^6=
		\left [\begin{matrix}
			B_{aa}& -B_{ab} &-B_{ac}\\
			-B_{ba}& B_{bb} &B_{bc}\\
			-B_{ca}& B_{cb} &B_{cc}\\
		\end{matrix} \right ].
  \label{Eq.S5}
\end{equation}
The matrix element of the six Ce sites are summed
\begin{equation}
		\begin{aligned}
			\mathbb{A}(\textbf{q})=&4e^{iq_cd_1}\times\left [\begin{matrix}
				A_{aa}c_ac_b& -A_{ab}s_as_b &iA_{ac}s_ac_b\\
				-A_{ba}s_as_b& A_{bb}c_ac_b & iA_{bc}c_as_b\\
				iA_{ca}s_ac_b& iA_{cb}c_as_b &A_{cc}c_ac_b\\
			\end{matrix} \right ]+2e^{iq_cd_5}\times\left [\begin{matrix}
				B_{aa}c_a& iB_{ab}s_a &iB_{ac}s_a\\
				iB_{ba}s_a& B_{bb}c_a &B_{bc}c_a\\
				iB_{ca}s_a&B_{cb}c_a &B_{cc}c_a\\
			\end{matrix} \right ],
		\end{aligned}
  \label{Eq.S6}
\end{equation}
where $c_a=\cos(q_aa/2)$, $c_b=\cos(q_bb/2)$, $s_a=\sin(q_aa/2)$, $s_b=\sin(q_bb/2)$ and $d_1=0.085c$, $d_5=0.165c$. $\mathbb{A}(\textbf{-q})$ is calculated in a similar way. Putting together,  the form factor can be derived as
\begin{equation}
		\begin{aligned} F_{aa}^{\mathbf{H} \parallel \mathbf{c}} (\mathbf{q})=&16[A_{aa}^2c_a^2c_b^2+A_{ba}^2s_a^2s_b^2]\\
			&+16[A_{aa}B_{aa}c_a^2c_b\cos(q_c(d_1-d_5))-A_{ba}B_{ba}s_a^2s_b\sin(q_c(d_1-d_5))]\\
			&+4(B_{aa}^2c_a^2+B_{ba}^2s_a^2).
		\end{aligned}
  \label{Eq.S7}
\end{equation}
 Due to the tetragonal crystal symmetry, $B_{ab}=0$ and we make the approximation $A_{aa}\approx A_{ab}$. The intensity of the transferred hyperfine field is strongly dependent on distance, and the distance of Ce1 to Al is approximately equal to Ce5 to Al. To simplify the calculations, we set $A_{aa}/B_{aa} \sim 1$. The distribution of $F_{aa}^{\mathbf{H} \parallel \mathbf{c}}(\mathbf{q})$ in momentum space was shown in Fig. 5, which has been renormalized by $F_{aa}^{\mathbf{H} \parallel \mathbf{c}}(\mathbf{0})$. The normalized hyperfine form factor decays from the
center of Brillouin zone, and retains a large value 0.558 at $\mathbf{q_m}$, demonstrating that spin fluctuations caused by
nesting between the Weyl nodes at ($\pm$0.2, $\pm$0.2, 0) would not be totally filtered out and thus can contribute substantially to $^{27}$Al $1/T_1T$.

\end{document}